\documentclass[11pt]{llncs}
\usepackage{wg06}
\usepackage[vlined, algoruled]{algorithm2e}

\newcommand{\mC}{\mathcal{C}}  
 \newcommand{\mF}{\EuScript{F}}
\newcommand{\mG}{\EuScript{G}}

 \newcommand{\mP}{\EuScript{P}}

 \newcommand{\mZ}{\mathcal{Z}}

\usepackage{fullpage}
\newcommand{\euss}{\EuScript S}

\newcommand{\et}{~\wedge~}
\newcommand{\impl}{~\Rightarrow~}
\newcommand{\equi}{~\Leftrightarrow~}
\newcommand{\brochette}{\cite{HPV99,McCS99,DGMcC01,McCS94,CH94,CHM02,HMP04,M03}\xspace}
\begin{document}
\title{Homogeneity vs. Adjacency: generalising some graph decomposition algorithms}
\titlerunning{Homogeneity vs. Adjacency}
\author{
B.-M. Bui Xuan\inst{1}
\and
M. Habib\inst{2}
\and
V. Limouzy\inst{2}
\and
F. de Montgolfier\inst{2}
}
\institute{\textsc{LIRMM}, Universit\'e Montpellier 2, France. \texttt{buixuan@lirmm.fr} \and 
\textsc{LIAFA}, Universit\'e Paris 7, France.
\texttt{\{habib,limouzy,fm\}@liafa.jussieu.fr}}
\maketitle              

\begin{abstract}
In this paper, a new general decomposition theory inspired from modular
graph decomposition is presented.
Our main result shows that, within this general theory, most of the nice
algorithmic tools developed for modular decomposition
are still efficient.

This theory not only unifies the usual modular decomposition
generalisations such as modular decomposition of directed graphs or decomposition of 2-structures, but
also star cutsets and bimodular decomposition.  Our general
framework provides a decomposition algorithm which improves the best known 
algorithms for bimodular decomposition.
\end{abstract}
\section{Introduction}\label{sec_intro}
Several combinatorial algorithms are based on partition refinement techniques \cite{MR917035}. Graph algorithms make an intensive use of vertex splitting,  the action of partitioning classes between neighbours and non-neighbours of a vertex. For instance, all known
linear-time modular decomposition algorithms \brochette
use this technique.

In bioinformatics also, the distinction of a set by a element, called a splitter, seems to
play an important role, as for example in the nice algorithm of
\cite{UY00}, which computes the set of common intervals of two
permutations.

In this paper we investigate an abstract notion of splitters and propose a dual formalism based on the concept of homogeneity.
Our aim is a better understanding of the existing modular decomposition algorithms by characterising the algebraic properties on which they are based.
Our main result  is that, within this general theory, most of the nice
algorithmic tools developed for modular decomposition \brochette are still efficient.

This theory not only unifies the usual modular decomposition
generalisations such as modular decomposition of directed graphs
\cite{McCFM05} or decomposition of 2-structures \cite{ER90}, but also
allows to handle star cutsets, and the bimodular decomposition
\cite{WG04}.  Notice that our general framework provides a
decomposition algorithm which improves the best known algorithm for
bimodular decomposition.

The paper is structured as follows: first we detail this new combinatorial decomposition theory,
then we present  a general algorithmic framework, and we finish by listing some interesting applications.


\section{Homogeneity, a new viewpoint}\label{sec_def}
Throughout this section $V$ is a finite set. The family of all subsets of $V$ is denoted $\mP(V)$. An \emph{reflectless}
triple is $(x,y,z)\subseteq V^3$ with $x\ne y$ and $x\ne z$. Reflectless
triples will be denoted $(x|yz)$ instead of $(x,y,z)$ since the first element does not play the same role.
Let $H$ be a relation over the reflectless triples of $V$. Given $s\in
V$, the relation $H_s$ is a binary relation on $V$ defined as $H_s(x,y)$ if and
only if $H(s|xy)$.

\begin{definition}[Homogeneous relation]
$H$ is a \emph{homogeneous relation on $V$} if, for all $s\in V$, $H_s$ is an equivalence relation on $V\setminus\{s\}$:\\
- (Symmetry):  $\forall \ s,x,y\in V,~H(s|xy)\equi H(s|yx)$.\\
- (Reflexivity): $\forall \ s,x\in V,~s\neq x\impl H(s|xx)$.\\
- (Transitivity): $\forall~s,x,y,z\in V,~H(s|xy)\et H(s|yz)\impl H(s|xz).$\\
\end{definition}

\begin{definition}[Homogeneous sets]\label{def1}
Let $H$ be a homogeneous relation.
$X\subseteq V$ is \emph{homogeneous with respect to an element $s\notin X$} if $H(s|xy)$ for all $x,y\in X$.
If $X$ is not homogeneous w.r.t. $s$ then $s$
\emph{distinguishes} $X$, or is a \emph{splitter} of $X$. Let $\euss_X$ be the set of all elements distinguishing $X$ and $s(X)=|\euss_X|$.

 $M\subseteq V$ is a \emph{homogeneous set} if $M\neq\emptyset$ and for all $x$ not in $M$, $M$ is
homogeneous w.r.t. $x$. In other words $s(M)=0$.  The family of homogeneous sets for a
homogeneous relation $H$ on $V$ is denoted $\mF_H$ or $\mF$ if not
ambiguous.
\end{definition}

\remark  From the definition it is obvious that, given a homogeneous set $M$, if $\neg H(s|xy)$ for some $x,y\in M$ then $s\in M$.

Two sets $A$ and $B$   \emph{overlap} if $A\cap B$, $A\setminus B$ and $B\setminus A$ are all nonempty. It is denoted $A\chev B$.
The \emph{symmetric difference} of two sets $A$ and $B$, denoted $A\Delta B$, is $(A\setminus B)\cup (B\setminus A)$.
Let us now enumerate some properties of the homogeneous relations and sets.

\begin{proposition}\label{prodeux}
If $s$ distinguishes $X$ then $\neg H(s|xy)$ for some $x,y\in X$.
\end{proposition}

Thanks to this proposition, it is exactly equivalent to define the
homogeneity relations as ternary relations on reflectless triples -- as
presented here -- or as relations between elements and subsets of $V$. A
homogeneous relation $\tilde{H}(s|X)$ for $X\subseteq V$ and $s\in
V\setminus X$ simply is a transitive relation: if $A\chev B$ and
$\tilde{H}(s|A)$ and $\tilde{H}(s|B)$ then $\tilde{H}(s|A\cup B)$. The
relation $\tilde{H}$ has the same properties than $H$ and the
homogeneous sets are the same.
We find the ternary relation much simpler.

\begin{proposition}\label{prop1}
For all $A, B\in \mF$ if $A\chev B$ then $(A\cap B)\in \mF$ and $(A\cup B)\in \mF$.
\end{proposition}

This property is called \emph{closure under intersection and union}. It is easy to check and can be used to prove:

\begin{proposition}[Lattice structure]
Let $H$ be a homogeneous relation on $V$ and $\mF'_H=\mF_H\cup\{\emptyset\}$.
$(\mF'_H,\subseteq)$ is a lattice.
\end{proposition}
\begin{proof}
Since $\emptyset\in\mF'_H$, and thanks to Proposition~\ref{prop1}, the
intersection of any two elements of $\mF'_H$ belong to $\mF'_H$.  It is
the infimum of two sets, since any member of $\mF'_H$ contained in both $A$
and $B$ is contained in $A\cap B$.  Let us consider the family $\mG$
of all set of $\mF'_H$ containing both $A$ and $B$. It is nonempty ($V$ is a member). Since $\mF'_H$ is closed under intersection, $\mG$
admits a unique smallest (w.r.t. inclusion) element, the intersection of
all its members, that is the supremum of $A$ and $B$.  \qed
\end{proof}

This lattice is a sublattice of the boolean lattice (hypercube) on $V$.  Moreover, if we consider $A \in \mF$ such that $|A| \geq 1$, $\mF(A)= \{\mF \in \mF_H~and~\mF \supseteq A\}$ then 
$(\mF(A),\subseteq)$ is a distributive lattice. Let us now define some useful types of homogeneous relations.
 
\begin{definition}
A homogeneous relation $H$ fulfills
\begin{itemize}
\item {\bf \emph{A1}} if  $\forall \ x,y,z\in V$,
$ H(x|yz)\et H(y|xz)\impl H(z|xy).$
\item {\bf \emph{A2}} if $\forall \ s,t,x,y\in V$,
$ H(x|st)\et H(y|st)\et H(t|xy)\impl H(s|xy).$
\item {\bf \emph{A3}} if $\forall \ s,t,x,y\in V$,
$ H(x|st)\et H(y|st)\et H(t|sx)\et H(t|sy)\impl H(s|xy).$
\item {\bf \emph{A4}} if $\forall \ x,y,z\in V$,
$ \neg H(x|yz)\et \neg H(y|xz)\impl H(z|xy).$
\end{itemize}
\end{definition}

\begin{proposition}[Quotient]
Let $H$ be a homogeneous relation.
Then, $H$ satisfies  {\bf  \emph{A2}} if and only if for all homogeneous set $M$, for all $x,y\in M$ and $s,t\in V\setminus M$,
$ H(x|st)\equi H(y|st)$.
\end{proposition}
It is a simple rewriting of {\bf \emph{A2}} but it enlightens that,
from a homogeneous set, one can pick a \emph{representative
element}.
Indeed, elements in a homogeneous set $M$ uniformly perceive a set $X$ not intersecting $M$: if one element of
$M$ distinguishes $X$ then so do all.  This allows to shrink a
homogeneous set $M$ into a single element, the quotient by $M$.

\medskip

Given $X\subseteq V$ one can define the induced relation $H[X]$ as $H$ restricted to reflectless triples of $X^3$. If $X$ is a homogeneous set we have the following nice property:

\begin{proposition}[Restriction]
Let $H$ be a homogeneous relation, $M$ a homogeneous set and $M'\subseteq M$.
$M'\in \mF_{H[M]}\ \equi \  M'\in \mF_H$.
\end{proposition}
 
Recursivity can therefore be used when dealing with homogeneous
sets. Notice that the proposition is not always true if $M$ is not a homogeneous
set. The Quotient and Restriction properties were used first with
modular decomposition and are useful for algorithmic \cite{MR84}.

\section{Submodularity of homogeneous relations}\label{sec_prop}
\begin{definition}
A set function $\mu:~\mP(V)\rightarrow\mathbb{R}$ is submodular if and only if  for all $X,Y\subseteq V$ $\mu(X)+\mu(Y)\geq\mu(X\cup Y)+\mu(X\cap Y)$ (see e.g. \cite{F91}).
\end{definition}

\begin{theorem}[Submodularity]\label{theo_submod}
Let $H$ be a homogeneous relation. The function $s$ counting the number of splitters (defined in Definition~\ref{def1}) is submodular.
\end{theorem}

\begin{proof}
Let us for convenience define $s(\emptyset)=-|V|$.  It suffices to prove $s(X)+s(Y)\geq s(X\cup Y)+s(X\cap Y)$ for all overlapping $X,Y\subseteq V$. So let $X,Y\subseteq V$ be two overlapping sets.
Obviously, ${\EuScript S}_{X\cap Y}=\left({\EuScript S}_{X\cap Y}\setminus Y,{\EuScript S}_{X\cap Y}\cap Y\right).$

As an element distinguishing $X$ does not belong to $X$,
 the partition ${\EuScript S}_{X\cup Y}=({\EuScript S}_{X\cup Y}\setminus {\EuScript S}_X,{\EuScript S}_{X\cup Y}\cap{\EuScript S}_X)$ can be reduced to
${\EuScript S}_{X\cup Y}=({\EuScript S}_{X\cup Y}\setminus {\EuScript S}_X,$ ${\EuScript S}_X\setminus (X\cup Y))$.
Similarly,
${\EuScript S}_Y=\left({\EuScript S}_Y\setminus {\EuScript S}_{X\cap Y},{\EuScript S}_{X\cap Y}\setminus Y\right).$
Finally,
${\EuScript S}_X=({\EuScript S}_X\setminus Y,({\EuScript S}_X\cap Y)\setminus {\EuScript S}_{X\cap Y},({\EuScript S}_X\cap Y)\cap{\EuScript S}_{X\cap Y})$
can be reduced to
${\EuScript S}_X=({\EuScript S}_X\setminus(X\cup Y),({\EuScript S}_X\cap Y)\setminus {\EuScript S}_{X\cap Y},{\EuScript S}_{X\cap Y}\cap Y).$
Hence, $|{\EuScript S}_X|+|{\EuScript S}_Y|-|{\EuScript S}_{X\cup Y}|-|{\EuScript S}_{X\cap Y}|=|({\EuScript S}_X\cap Y)\setminus {\EuScript S}_{X\cap Y}|+|{\EuScript S}_Y\setminus {\EuScript S}_{X\cap Y}|-|{\EuScript S}_{X\cup Y}\setminus {\EuScript S}_X|.$

To achieve proving the theorem, we prove that ${\EuScript S}_{X\cup Y}\setminus {\EuScript S}_X\subseteq{\EuScript S}_Y\setminus {\EuScript S}_{X\cap Y}$.
Indeed, let $z\in{\EuScript S}_{X\cup Y}\setminus{\EuScript S}_X$.
Then, $z\notin X\cup Y$ and for all $x,y\in X$, we have $ H(z|xy)$.

Now, suppose that $z\notin\euss_Y$.
Since $z$ is not in $X\cup Y$, we have $ H(z|xy)$ for all $x,y\in Y$.
Furthermore, as $X$ and $Y$ overlap and thanks to the transitivity of $H$, we have $z\notin\euss_{X\cup Y}$, which is a contradiction.

Finally, supposing $z\in\euss_{X\cap Y}$ would imply $z\in\euss_X$.
\qed
\end{proof}

Submodular functions are  combinatorial objects with powerful
potential (see e.g.~\cite{F91}).  Theorem~\ref{theo_submod} enables
the application of this theory to homogeneous relations.
In~\cite{UY00}, T. Uno and M. Yagiura gave a (restricted) version of
this theorem, and constructed a very nice algorithm computing the
common intervals of a set of permutations.  It would be interesting to
generalise this approach to any homogeneous relation, as done
in~\cite{BHP06} for modular decomposition.

\section{Strong homogeneous sets and Primality}
In a family $\mF$ of subsets of $V$, a subset is \emph{strong} if it
overlaps no other subset of $\mF$. The other subsets are
\emph{weak}.
Let us suppose the family $\mF$ contains $V$ and the singletons $\{v\}$ for every element.
Then $V$ and $\{v\}_{v\in V}$ form the \emph{trivial} strong subsets.
The set inclusion orders the strong
subsets into a tree.  This is a quick proof that there are at most
$2|V|-1$ strong subsets (and at most $|V|-2$ nontrivial ones), as the
tree has no internal node of degree 1.

The \emph{parent} of a (possibly weak) subset $M$ is the smallest
strong subset $M_P$ properly containing $M$, and $M$ is said to be a
\emph{child} of $M_P$. If $M$ is strong, $M_P$ is by definition its
parent in the inclusion tree.

An \emph{overlap class} is an equivalence class of the transitive
closure of the overlap relation $\chev$ on $\mF$. The \emph{support}
of an overlap class $\mC=\{C_1,...C_k\}$ is $C_1\cup...\cup C_k$. $A$
is an \emph{atom} of the overlap class if it is included in at least
one subset $C_i$, and it does not overlap any subset of the class, and
is maximal for these properties. All the atoms of a class form a
partition of its support, the coarsest partition compatible with the
class. An overlap class is \emph{trivial} if it contains only one
subset; it is then clearly a strong one.

A strong subset is \emph{prime} if all its children are strong, and
\emph{decomposable} otherwise. It is a classical result of set theory that
\begin{lemma}\label{lemover}
If $\mF$ is a family closed under union of overlapping sets, then
there is an one-to-one correspondence between the nontrivial overlap
classes of $\mF$ and the decomposable strong subsets of $\mF$. More
precisely, the overlap class $\mC$ associated with a decomposable
subset $D$ is simply the set of weak children of $D$, and the support
of $\mC$ is $D$.
\end{lemma}

The overlap class associated with a decomposable node is simply the set
of weak children of this node.
Of course we apply all these notions on homogeneous set families.  On partitive families, the
strong homogeneous sets plays a very important role since they are
exactly a coding, in $O(|V|)$ space, of the possibly $2^{|V|}$ subsets
of the family (see the upcoming section).

\begin{theorem}\label{thfab}
Let $H$ be a homogeneous relation and $\mZ$ be the family of
homogeneous sets containing $x$ but not $y$, and maximal for this
property, for all $x$ and $y$.  The strong homogeneous sets of
$H$ are exactly the supports and atoms of all overlap classes of $\mZ$ 
\end{theorem}
\begin{proof}
First, remark that, thanks to the closure under union of overlapping
sets, the supports and atoms of every overlap class of $\mZ$ are
strong homogeneous sets. Lemma~\ref{lemover} tells they can not be
overlapped by an element of $\mZ$ and if one, $A$ is overlapped by an
homogeneous set $B\notin \mZ$ then for $x\in A\setminus B$, the
maximal homogeneous set containing $y$ but not $x$ overlaps $A$, a
contradiction. So the family of supports and atoms is included in the
family of strong homogeneous sets. Conversely, let us prove that if $M$
is a strong homogeneous set then it is the support or an atom of some
overlap class. We shall distinguish four cases. Let $M_P$ be the
strong parent of $M$ (for $M\ne V$).

\begin{enumerate}
\item $M$ is trivial ($V$ or $\{v\}$). There is no problem.
\item $M$ is decomposable. It has $k$ strong children $M_1..M_k$. Let
us pick an element $x_i$ in each $M_i$. Then for all $i$ and $j$ we
consider the maximal homogeneous set containing $x_i$ but not $x_j$. They
form an overlap class of $\mZ$. Its support is $M$, thanks to
Lemma~\ref{lemover}
\item $M$ is prime and $M_P$ is prime. Then for all $x\in M$ and all
$y\in M_P\setminus M$ $M$ is the maximal homogeneous set containing
$x$ but not $y$. As it is strong, it belongs to a trivial overlap
class and is equals to its support.
\item $M$ is prime and $M_P$ is decomposable.  Then for all $x\in
M_P\setminus M$ then $M$ is included in some maximal homogeneous set $M_x$ not containing $x$ (the one that contains the vertices of $M$).
Let us consider the intersection $I$ of all subsets of
$\{M_x \ | \ x\in M_P\setminus M\}$. It is an atom of the overlap
class associated with $M_P$ and thus is strong. As $M$ is a children
of $M_P$, $I=M$.\qed
\end{enumerate}

\end{proof}

This theorem leads to a $O(|V|^3)$-time algorithm in Section~\ref{sectfort}.

\section{Partitive families of homogeneous sets}
A generalisation of modular decomposition, known from \cite{CHM81}, less
general than homogeneous relations but more powerful, is the
\emph{partitives families}.


\begin{definition}
A family $\mF\subseteq \mP(V)$ is \emph{weakly partitive} if it contains $V$ and the singletons $\{v\}$ for all  $v\in V$, and is closed under union, intersection and difference of overlapping subsets, i. e.
$$A\in \mF_H \et B \in \mF_H \et A \chev B \impl A\cap B\in \mF_H \et  A\cup B\in \mF_H \et   A\setminus B\in \mF_H$$  
Furthermore a weakly partitive family $\mF$ is \emph{partitive} if it is also closed under symmetric difference:
$$A\in \mF_H \et B \in \mF_H \et A \chev B \impl A\Delta B\in \mF_H$$  
\end{definition}

As mentionned before, strong subsets of a weakly partitive family
$\mF$ can be ordered by inclusion to a tree. Let us define three types
of strong subsets, i.e. three types of nodes of the tree:
\begin{itemize}
\item \emph{prime} nodes who have no weak children,
\item \emph{degenerate} nodes such that all union of strong children of the node belongs to ${\cal F}$,
\item\emph{linear} nodes such that there is an ordering of the strong children such that a union of children belongs to ${\cal F}$ if and only if they are consecutive in this ordering.
\end{itemize}

\begin{theorem}[\cite{CHM81}]
In a partitive family, there exists only prime and degenerate nodes.
In a weakly partitive family, there exists only prime and degenerate and linear nodes.
\end{theorem}
The strong subsets are therefore an $O(|V|)$ space encoding of the family: it is enough to type the nodes into complete, linear or prime, and to order the children of the linear nodes. All weak subsets can be output just by making simple combinations of the strong children of decomposable (complete or linear) nodes.
Now, the following properties state that the homogeneous relations are a proper generalisation of (weakly) partitives families.

\begin{proposition}\label{lemm_p_wp}
Let $H$ be a homogeneous relation.
If $H$ fulfills {\bf \emph{A1}}, or if $H$ fulfills {\bf \emph{A2}}, then  $H$ fulfills {\bf \emph{A3}}.
\end{proposition}
\remark
This proposition allows to classify the homogeneous relations. There exists homogeneous relations fulfilling {\bf \emph{Ai}} but not {\bf \emph{Aj}} for all $i$ and $j$ excepted the two implications of Proposition~\ref{lemm_p_wp}.

\begin{proposition}\label{prop_wp}
If a homogeneous relation $H$ fulfills {\bf \emph{A2}} or {\bf \emph{A3}}, then $\mF_H$ is a weakly partitive family.\\
If $H$ fulfills {\bf \emph{A1}}, then $\mF_H$ is a partitive family.
\end{proposition}
\begin{proof}
Let us suppose $A\in \mF_H$ and $B \in \mF_H$ and $A \chev B$. Thanks
to transitivity an element not in $A\cup B$ can not distinguish $A\cup
B$ (it would distinguish $A$ or $B$). As an element not in $A$ can not
distinguish $A$ and an element not in $B$ can not distinguish $B$, then no
element can distinguish $A\cap B$. For the same reason, only an
element of $A\cap B$ can distinguish $A\setminus B$ or $A\Delta B$.

If $s\in A\cap B$ distinguishes $A\setminus B$, then this set contains
$x$ and $y$ such that $\neg H(z|xy)$. But as $B\setminus A$ is
nonempty it contains $t$ and we have $H(x|st)$ and $H(y|st)$ and
$H(t|sx)$ and $H(t|sy)$ and $H(t|xy)$. Then both {\bf \emph{A2}} and
{\bf \emph{A3}} are violated.

Let us suppose {\bf \emph{A1}} holds. As {\bf \emph{A3}} also holds
$A\setminus B$ and $B\setminus A$ are homogeneous sets. If $z\in A\cap
B$ distinguishes $A\Delta B$, then there exists $x\in A$ and $y\in B$
such that $\neg H(z|xy)$. But since $H(x|yz)$ and $H(y|xz)$ {\bf
\emph{A1}} is contradicted.
\qed
\end{proof}

\section{Homogeneous Set Algorithms}\label{sectalgo}

In the following, we consider a fixed ground set $V$ and a homogeneous relation $H$ on $V$, that are the input of all algorithms described here. The input $H$ consists in $|V|$ partitions (the equivalence classes of $H_x$ for each $x$) and thus can be stored in $O(|V|^2)$ space, instead of the naive $O(|V|^3)$ space representation storing all triples.

\subsection{Smallest Homogeneous Set containing a subset}
Let $H$ be a homogeneous relation on $V$ and $S$ be a nonempty subset
of $V$. As $\mF_H$ is closed under intersection, there is an unique
smallest homogeneous set containing $S$, the intersection of all
homogeneous set containing $S$, denoted henceforth $SHS(S)$.
\begin{algorithm}[t!]
\caption{Smallest Homogeneous set containing $S$} \label{algoSHS} 
\dontprintsemicolon 
Let $x$ be an element of $S$, $M:=\{x\}$ and $F:=S\setminus\{x\}$\;
\While{$F$ is not empty}{
  pick an element $y$ in $F$ ; $F :=  F\setminus\{y\}$ ; $M :=  M\cup\{y\}$  \;
  \For{every element $z$}{
     \lIf{$H(z|x,y)$}
         {$F := F \cup \{z\}$\;}
}
output $M$ (now equals to $SHS(S)$)\; 
}
\end{algorithm}

\begin{theorem}
Algorithm \ref{algoSHS} computes $SHS(S)$ in $O(|V|.|SHS(S)|)= O(|V|^2)$ time. 
\end{theorem}
\begin{proof}
Time complexity is obvious as the \textbf{while} loop runs $|M|-1$
times and the \textbf{for} loop $|V|$ times. The algorithm maintains
the invariant that every splitter of $M$ is in $F$. When $M$ is
replaced by $M\cup\{y\}$, every element that distinguishes $M\cup\{y\}$
distinguishes $x$ from $y$, or already is in $F$. The algorithm ends
therefore on a homogeneous set that contains $S$, and thus we have
$SHS(S)\subseteq M$. If $M\ne SHS(S)$ let $v$ be the first element of
$M\setminus SHS(S)$ added to $F$ (eventually added to $M$). It
distinguished two elements $x$ and $y$ from $SHS(S)$, contradicting
its homogeneity. So $SHS(S)=M$.
\qed
\end{proof}

\subsection{Maximal Homogeneous Sets not containing an element}\label{sectmax}

\begin{proposition}
Let $H$ be a homogeneous relation on $V$ and $x\in V$ an element.  As
$\mF_H$ is closed under union of intersecting subsets, there is an
unique partition of $V\setminus\{x\}$ into $S_1...S_k$ such that every
$S_i$ is a homogeneous set of $\mF_H$ and is maximal w.r.t. inclusion in  $\mF_H$.
\end{proposition}
We call $MHS(x)\subset\mP(V)$ this partition of Maximal Homogeneous
Sets not containing $x$. We propose a partition refining algorithm \cite{MR917035}. It
is obvious that

\begin{lemma}
Every homogeneous set (especially the maximal ones) not contains $x$
is included in a $x$-class $H_x^i$ of $H$.
\end{lemma}

Therefore our algorithm starts with the partition $P=\{H_x^1..H_x^k\}$
of the $x$-classes of $H$. Then the partition is refined (classes are
splitted) using the following rule. Let $y$ be an element, called the
\emph{pivot}, and $C(y)$ the class of Partition $P$ containing $y$.

\medskip

\noindent\textbf{Rule.} \textsl{For a given pivot $y$, split every class of $P$, excepted $C(y)$, into $C\cap H_y^1$,...,$C\cap H_y^l$}

\medskip

Notice that a class is actually splitted in many new classes iff it is
distinguished by $y$.

\begin{lemma}\label{lemsplit}
Starting from the partition $P_0=\{H_x^1..H_x^k\}$, the application of
the refining rule (for any pivot in any order) until no class can
be actually splitted, produces $MHS(x)$
\end{lemma}
\begin{proof}
The refining process ends when no pivot can split a class, i.e when
every partition class is a homogeneous set. Let us suppose one of
these homogeneous sets $M$ is not maximal w.r.t. inclusion: it is
included in a homogeneous set $M'$, itself included in a $x$-class
$H_x^i$. Let us consider the pivot $y$ that first broke $M'$. It can
not be out of $M'$, as $M'$ is homogeneous, nor within $M'$, as a
pivot does not break its own class. But $M'$ was broken, contradiction.
\qed
\end{proof}

Let us now implement this lemma into an efficient algorithm.  The idea
for saving time is that, after a class is splitted by $y$, it has not to
be examined one more time but, if the former class $C$ containing
$y$ is later splitted into new classes $C_1..C_a$. W.l.o.g, suppose $y\in C_1$.
Then only the new classes $C_2..C_a$ must be examined when $y$ is
the pivot again. Every element is thus examined once for every pivot,
leading to an $O(|V|^2)$ time complexity. This is implemented in
Algorithm~\ref{algoMHS}.

We implement this idea using \emph{groups}. The partition is refined
from $P_0$ into $P_1$ then $P_2$ and so on. The \emph{group} of an element
 of $P_i$ is its class in $P_{i-1}$. If the classes are
implemented using a linked list, and if the classes are only splitted
into new classes that follow consecutively in the linked list, then
the group boundaries are simply markers in the linked list.

 A classical trick of partition refining \cite{MR917035,HPV99} is using a \emph{refining set}
$R\subseteq V$. Every class $C$ of $P$ can be splitted into $C\cap R$ and
$C\setminus R$ in $O(|R|)$ time only: every element of $R$ is moved
from its old class $C_i$ to its new class $C_i'$, the successor of
$C_i$ in the linked list. A flag in the data structure of $C_i$ indicates wether
its successor is $C_i'$ or not. This allows to create $C_i'$ if it does
not already exist. $C_i'$ is $C_i\cap R$ while the remaining elements
of $C_i$ are $C_i\setminus R$.  All flags are reset by a second scan of
$R$, that also allows to remove empty classes from the partition
linked list (classes that were included in $R$).

And at least, for a set $Z$ and $y\notin Z$, $Z$ can be partitioned
according to the $y$-classes in $O(|Z|)$ time. If there are $k$
$y$-classes, an array of $k$ linked list is used and each element of
$Z$ is appended to the proper list. A stack of nonempty list allow to
collect and reset them in $O(|Z|)$ time.

\begin{algorithm}[ht!]
\caption{Maximal Homogeneous Sets not containing  $x$} \label{algoMHS} 
\dontprintsemicolon 
\For{ every group $G$}{
\For{ every class $C$ of $G$}{
Compute the set $Z$ of elements of in $G$ but not in $C$\;
\For{ every element $y$ of $C$}{
Partition $Z$ according to the $y$-classes\;
Add each partition set to the refining sets pool
}}}
Set the group boundaries to the classes boundaries (from $P_{i-1}$ to $P_i$)\;
\For{ each refining set $R$ of the pool}{
Remove $R$ from the pool and then refine  $P_i$ using $R$
}
\end{algorithm}

\begin{theorem}
Algorithm \ref{algoMHS} computes $MHS(x)$ in $O(|V|^2)$ time. 
\end{theorem}
\begin{proof}
For the correctness proof, one just has to check that the algorithm
implements correctly Lemma~\ref{lemsplit}. For the time
complexity issues, notice that, for each pivot $y$, an element $z$ is placed in $Z$
only once. As partitioning $Z$ into to $y$-classes, and then refining
using all refining sets generated by $y$, takes $O(|Z|)$ time. Hence the algorithm takes
$O(|V|^2)$ time.  \qed
\end{proof}

\subsection{Testing if a homogeneity relation is trivial}
A homogeneous relation $H$ on $V$ is \emph{trivial} if $\mF_H$ contains only $V$ and the singletons.

\begin{theorem}
Let $H$ be a homogeneous relation on $V$ and $S$ be a nonempty subset
of $V$. One can test in $O(|V|^2)$ time if $H$ is trivial.
\end{theorem}
\begin{proof}
If $|V|<2$ the answer is yes. Otherwise let $x$ and $y$ be two
elements of $V$. In $O(|V|^2)$ time, Algorithm~\ref{algoMHS} outputs
the maximal homogeneous sets not containing $x$. If one of them is
nontrivial the answer is no. Otherwise all nontrivial homogeneous sets
contain $x$.  In $O(|V|^2)$ time, Algorithm~\ref{algoMHS} outputs
the maximal homogeneous sets not containing $y$. If one of them is
nontrivial the answer is no. Otherwise all nontrivial homogeneous sets
contain $x$ and $y$. Then  Algorithm~\ref{algoSHS} is used with $S=\{x,y\}$, in $O(|V|^2)$ time. The answer is yes iff $SHS(\{x,y\})=V$.
\qed
\end{proof}

\subsection{Strong modules of a homogeneous relation}\label{sectfort}
Theorem~\ref{thfab} straightforwardly leads to an algorithm:
\begin{theorem}
The strong homogeneous subsets of a homogeneous relation $H$ on $V$ can be computed in $O(|V|^3)$ time.
\end{theorem}
\begin{proof}
First compute $MHS(x)$ for all $x\in V$. All these sets together
exactly form the family $\mZ$ defined in Theorem~\ref{thfab}. It can
be done in $O(|V|^3)$ time using Algorithm~\ref{algoMHS} $|V|$
times. The size of this family (sum of the cardinals of every subsets)
is $O(|V|^2)$ since they form $|V|$ partitions. Using Dahlhaus
algorithm \cite{D00} the overlap components can be found in time linear
in the size of the family, thus $O(|V|^2)$. According to
Lemma~\ref{lemover} there are at most $|V|$ nontrivial overlap
classes. For each class it is easy to compute its support, and in
$O(|V|^2)$ time easy to compute its atoms (each subset of the class is
used as pivot, in a partition refinement of the support). And after
all the $O(|V|^2)$ supports and atoms output must be sorted by
inclusion order into the inclusion tree of the strong homogeneous sets
(removing many duplicates), an easy task in $O(|V|^3)$ time.
\qed
\end{proof}
Notice that, if the homogeneous relation defines a weakly partitive
family, then the quotient property applies and helps a lot.  The
algorithm scheme of \cite{DGMcC01}, that can be implemented in
$O(n+m)$ for graphs, could be implemented in $O(|V|^2)$ time for an
homogeneous relation. The approach is to compute $MHS(x)$ then, using
the quotient relation, to compute all strong homogeneous sets
containing $x$. That gives the left branch of the decomposition
tree. Then the algorithm is recursively launched.  The amortised
complexity analysis of Section~\ref{sectmax} can be used: as a class
is not splitted when the recursive process begins, the whole algorithm
takes $O(|V|^2)$. But our proof needs Axiom~{\bf \emph{A4}}. It holds
for graphs, but not for directed graphs nor 2-structures, and seems
very specific.

\section{Applications}
Let us examine in the sequel some of the applications of this Homogeneity theory.

\subsection{Modular decomposition}
In a graph, the homogeneous relation $H(x|yz)$ is true when $x$
``sees'' $y$ and $y$ in the same way. In undirected graphs, this means
that either there are two edges $(xy)$ and $(xz)$, or no edge between
$x$ and the two other vertices. In directed graphs, this means that
there are zero or two incoming arcs between $x$ and the two other
vertices, and zero or two out-coming arcs between $x$ and the two
other vertices. The homogeneous sets are then called \emph{modules}. The notion of modules also extends to 2-structures~\cite{ER90}. A 2-structure is a complete edge-coloured graph and $H(x|yz)$ is true when edges $(xy)$ and $(xz)$ have the same colour. 

\begin{proposition}
\begin{itemize}
\item The homogeneous relation  of a undirected graph fulfills {\bf \emph{A1}}, {\bf \emph{A2}}, {\bf \emph{A3}} and {\bf \emph{A4}}
\item The homogeneous relation  of a directed graph fulfills {\bf \emph{A2}} and {\bf \emph{A3}} 
\item The homogeneous relation  of a 2-structure fulfills {\bf \emph{A1}} and {\bf \emph{A2}} and {\bf \emph{A3}} and {\bf \emph{A4}}
\end{itemize}
\end{proposition}
The modules of a undirected graph and of a 2-structure thus form a
partitive family, while the modules of a directed graph just form a
weakly partitive family. All know properties of modular decomposition
\cite{MR84} can be derived from this result. An $O(n^3)$ modular decomposition algorithm can also be derived from Section~\ref{sectfort} algorithm, but it is less efficient than the existing algorithms \brochette.

\subsection{Other graph relations}
In a graph we can consider different homogeneous relations, for instance the relation \emph{``there exists a path from 
vertex $x$ to vertex $y$ avoiding the vertex $s$"}, or a more general relation \emph{``there exists a path from $x$ to 
$y$ avoiding the neighbourhood of $s$"}. It is easy to see that these
two relations fulfill the basic axioms (symmetry, reflexivity and transitivity). In the first case, the strong homogeneous sets form a partition (into the 2-vertex-connected components, minus the articulation points). The second relation is related to decomposition into star cutsets.

Another interesting relation is $D_k(s|xy)$ if
$d(s,x)\le k$ and $d(s,y) \le k$, where $d(x,y)$ denotes the distance  between $x$ and $y$. The case $k=1$ 
corresponds to modular decomposition. It is worth investigating the general problem.

\subsection{Bimodular decomposition}
Let $G=(B,W,E)$ be a bipartite graph where $B$ contains the black vertices
and $W$ contains the white vertices. A \emph{bimodule} is a subset of vertices
$M\subset (B\cup W)$ such that no black vertex $b$ not in $M$
distinguishes two white vertices of $M$ (there must be either no or
all possible edges between $b$ and the white vertices of $M$) and
conversely no white vertex not in $M$ distinguishes two black vertices
of $M$. In \cite{WG04} is defined the \emph{bimodular decomposition}
of a bipartite graph. It is stated that, although the family of
bimodules is not even partitive, then the strong bimodules are an
optimal encoding of the family. Indeed, the inclusion tree of strong
bimodules, plus some $O(n^2)$ pointers (easy to add, given the graph and
the inclusion tree), are enough to store and output the (potentially
exponential) family of bimodules, and to test if a set is a bimodule, and
allow to solve in polynomial time some NP-complete problems, when the
degree of the nodes of the tree is bounded \cite{Loz02}. \cite{WG04}
give an $O(mn^3)$ time algorithm that compute the strong bimodules
given the graph. But given the graph the homogeneous relation can be
computed in $O(n ^3)$ time by testing all reflectless triples, and
then using the algorithm of Section~\ref{sectfort} the strong
bimodules can be output in $O(n ^3)$ time, improving the previous time
bound.

\section{Conclusion}
We hope that  this homogeneity theory will have many other 
applications and will be useful to decompose automata~\cite{AM03} and boolean functions~\cite{B05}.
Obviously, the algorithmic framework presented here can be optimised in each particular application, as it 
can be done for modular decomposition \brochette.  We think the homogeneity concept is a very 
general idea.


\bibliography{wg06}
\bibliographystyle{plain}


\end{document}